# Color intensity projections provides a fast, simple and robust method of summarizing the grayscale images from a renal scan in a single color image

Keith S Cover[1], Sandra A. Srbljin[2], Arthur van Lingen[2]

[1]Department of Physics and Medical Technology, [2]Department of Nuclear Medicine and PET Research,
VU University Medical Center, Amsterdam,

Corresponding author:
Keith S Cover, PhD                                       Email: Keith@kscover.ca
Department of Physics and Medical Technology             Tel:    31 20 444-0677
VU University Medical Center                             Fax:    31 20 444-4816
Postbus 7057
1007 MB Amsterdam
The Netherlands

Abstract

To assess its usefulness, the peak version of color intensity projections (CIPs) was used to display a summary of the grayscale images composing a renogram as a single color image. **Method** For each pixel in a renogram, the time point with the maximum intensity was used to control the hue of the color of the corresponding pixel in the CIPs image. The hue ranged over red-yellow-green-light blue-blue with red representing the earliest time. **Results** For subjects with normal appearing kidneys, the injection site shows up in red, the kidneys in a red-yellow and the bladder in a green-blue. A late fill kidney typically appeared greener or bluer than a normal kidney indicating it reached its peak intensity at a later time point than normal. **Conclusions** Having the time and intensity information summarized in a single image promises to speed up the initial impression of patients by less experienced interpreters and should also provide a valuable training tool.

## INTRODUCTION

Radionuclide imaging has been used for many decades to measure the uptake and excretion of the kidneys (TaplinGV1956, MettlerFA2006). Images are acquired over a 20 to 30 minute period using a gamma camera after the injection of a radionuclide. Each image is the accumulation of activity over a 1 second to 15 second period with shorter windows at earlier times.

The basic display and analysis of the renograms has changed little over the years although modern computers has made the processing much quicker, simpler and more reliable. Typically, a series of the images is reviewed repeatedly. In addition, ROI's are often drawn around each of the kidneys, and an average time curve for each of the kidneys plotted and examined in detail. The ROI's may be drawn on a single image or on an average of a few adjacent images.

Color intensity projections (CIPs) are a simple technique for summarizing a series of grayscale images in a single color image (CoverKS2006, CoverKS2007A, CoverKS2007B). It has been applied to a wide variety of applications including 4DCT, angiography, perfusion MRI and astronomy.

If all the grayscale images are identical then the resulting CIPs will also be a grayscale image that is identical to the component grayscale images. However, if on a pixel by pixel basis there are changes over the images, the amount and timing of the changes will introduce specific colors into corresponding pixels the CIPs image. Depending on the complexity of the motion, a CIPs can often provide most, if not all, of the important information that would be gained from examining the individual grayscale images.

**MATERIALS AND METHODS**

A total of 80 images were acquired from each of four subjects after the injection of an $^{99m}$Tc-labeled agent. Each image was accumulated over a 15 second interval yielding a total of 20 minutes of images for each subject.

The peak version of the CIPs was found to be the most useful for renograms (CoverKS2007B). For each pixel in a renogram, the time point with the maximum intensity was found. The brightness and time point of this maximum intensity value was used to control the color of the corresponding pixel in the CIPs image. The brightness of the pixel was set to be the brightness of the maximum value. The timing of the maximum intensity value was used to select the hue of the pixel in the CIPs. The hue values used ranged over red-yellow-green-light blue-blue with the earliest times corresponding to red and the latest to blue.

Renograms of two subjects were used to demonstrate the application of CIPs. Both subjects were scanned because they had shown indications of renal dysfunction. The first subject was a 12 year old female while the second was a 3 year old male.

**RESULTS**

Figure 1 shows the renogram of the first subject displayed as a CIPs. The injection site is clearly visible in the right arm. The counts from the arm appear in the first image acquired. Thus all the pixels are the same hue of red indicating very early time. However, as expected, the pixels do vary in intensity. The isotopes rapid movement into the lungs is also clearly evidenced by the red pixels in the lungs.

The display of the two kidneys in the CIPs are particularly informative. The first impression is the two kidneys are largely symmetric with each other, an indication of normal kidney function. More careful examination of the two kidneys shows redder pixels in the lateral regions of the kidneys and some yellow and light green pixels in the medial regions indicating the isotope reached the lateral regions 4 to 8 minutes earlier than the medial regions. This is consistent with the normal passage of isotope through the kidneys.

Figure 1 also shows the renograms curves for ROI's of the left kidney and of the bladder. The curve for the left kidney shows a peak intensity at 3 to 4 minutes, again in line with a normal kidney. The renogram curve for the bladder is included to demonstrate a late peaking curve. The green-light blue-blue pixels of the bladder indicate a curve that plateaued between 10 and 20 minutes. Examination of the bladder curve shows this is the case.

Figure 2 shows the CIPs of the renogram for the second subject. It also includes curves for the right kidney and the lower region of the left kidney. Examination of the right kidney in the CIPs shows isotope peaking at 2 to 3 minutes, consistent with a healthy kidney. Again most of the lateral pixels are slightly redder than the medial pixels, indicating the expected motion of the isotope through the kidney.

The upper region of the left kidney shows a similar pattern to the right kidney. However, the lower region of the left kidney is primarily green, indicating a peak of intensity at about 10 minutes, abnormally late. The late peak is suggestive of a blockage.

**CONCLUSIONS**

The presentation of a renogram as CIPs provides a simple and intuitive way to summarize a renogram in a single color image. Presenting a renogram as a CIPs should be helpful in interpreting renograms for those with less experience especially for training purposes. The CIPs may also be useful diagnostically but more studies are warranted, especially into the influence of patient movement. While color intensity projections has only been applied to renograms in this publication, it well may be a useful display technique for a range of other nuclear medicine modalities.

**ACKNOWLEDGEMENTS**

Thanks to Prof. Otto S. Hoekstra assistance with this study. One of the authors (KSC) was funded by the Department of Medical Technology and Physics, VU University Medical Center, Amsterdam. The VU University Medical Center, where all the authors work, is pursuing a patent on color intensity projections and one of the authors is listed as an inventor (KSC).

**Figure Captions**

**FIGURE 1.** Renogram displayed as a CIPs for the first subject. The count curves for the left kidney (L) and the bladder (B) are displayed below the CIPs.

**FIGURE 2.** Renogram displayed as a CIPs for the second subject. The count curves for the right kidney (R) and the lower region of the left kidney (LL) are also displayed.

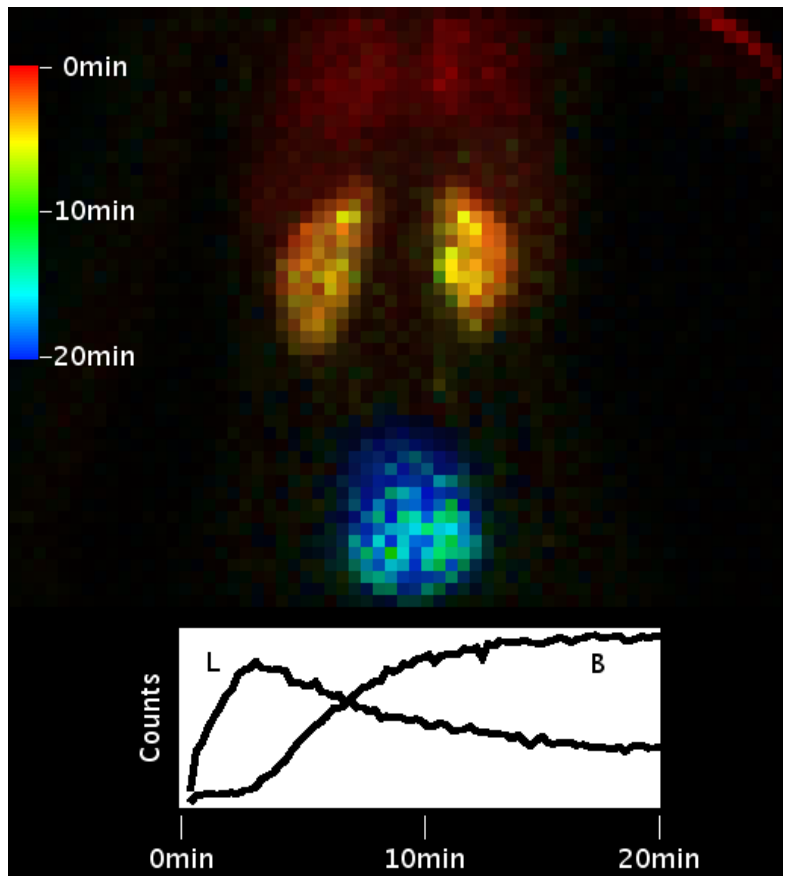
Fig 1

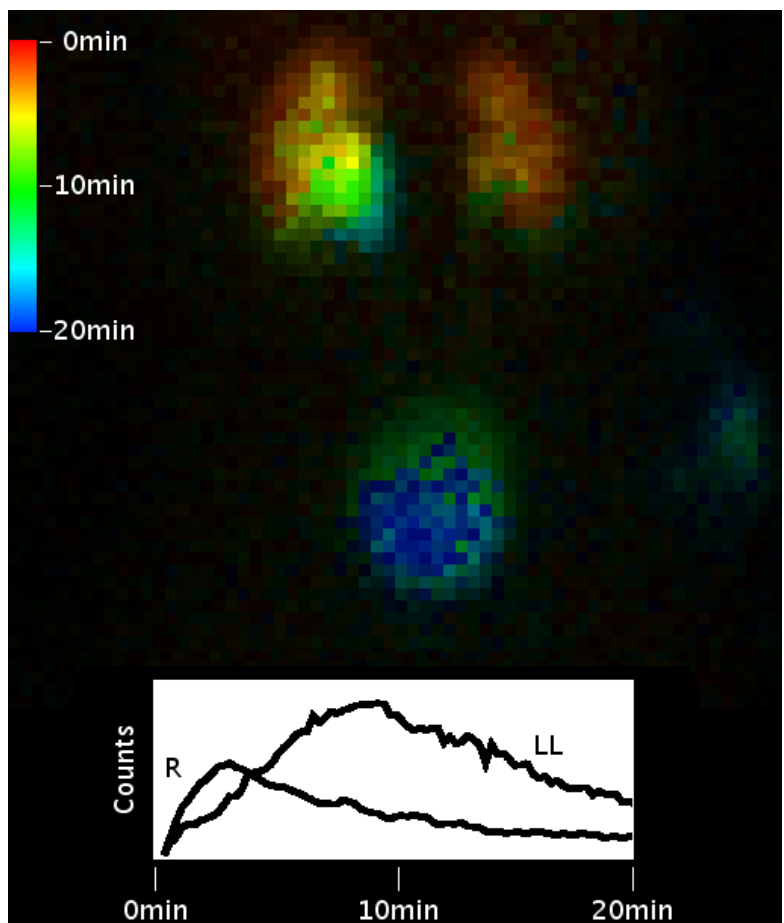
Fig 2